\documentclass[]{spie}  
\usepackage[]{graphicx}

\title{High Contrast L' Band Adaptive Optics Imaging to Detect Extrasolar Planets} 

\author{Ari Heinze, Phil Hinz, Suresh Sivanandam, Daniel Apai, and Michael Meyer\supit{a}
\skiplinehalf
\supit{a}All authors affiliated with Steward Observatory, University of Arizona, 933 N Cherry Ave, Tucson, AZ, 85721, USA;
}

\authorinfo{Ari Heinze is the corresponding author.  Email aheinze@as.arizona.edu}

\begin{document} 
  \maketitle 

\begin{abstract}
We are carrying out a survey to search for giant extrasolar planets
around nearby, moderate-age stars in the mid-infrared L' and M bands
(3.8 and 4.8 microns, respectively), using the Clio camera with the
adaptive optics system on the MMT telescope.  To date we have observed
7 stars, of a total 50 planned, including GJ 450 (distance
about 8.55pc, age about 1 billion years, no real companions detected), which
we use as our example here.  We report the methods we
use to obtain extremely high contrast imaging in L', and the performance
we have obtained.  We find that the rotation of a celestial object over time
with respect to a telescope tracking it with an altazimuth mount can be a powerful tool
for subtracting telescope-related stellar halo artifacts and 
detecting planets near bright stars.  We have carried out a thorough
Monte Carlo simulation demonstrating our
ability to detect planets as small as 6 Jupiter masses around
GJ 450.  The division of a science data set into two independent
parts, with companions required to be detected on both in order
to be recognized as real, played a crucial role in detecting companions in this simulation.
We mention also our discovery of a
previously unknown faint stellar companion to another of our
survey targets, HD 133002.  Followup is needed to confirm this as
a physical companion, and to determine its physical properties.
\end{abstract}



\section{INTRODUCTION}
\label{sect:intro}  

Slightly over a decade ago, the observational science of extrasolar
planetary systems began with the discovery of the planet 51 Pegasi B
by Michel Mayor and Didier Queloz.  Since that time nearly 200 additional
extrasolar planets have been discovered, mostly by using precise spectroscopic
radial velocity measurements to detect the reflex motion of the parent star in response to the
orbiting planet's gravity -- the same method used in the 51 Pegasi detection.
Direct imaging of extrasolar planets remains an elusive goal.

It is a very attractive goal, however.  A planet detected by direct
imaging could be investigated with spectroscopy and multi-band
photometry, and astrometric measurements over time would yield
its exact orbit.  Far more could be learned about such a planet
than about one that is only detected by the radial velocity method.
Also, the radial velocity method is most sensitive to planets
in fairly close-in orbits, unlike the giant planets in our own solar system.
Direct imaging is needed to find planets further out, and thus
systems more analagous to our own.

Immediately after their formation in the protoplanetary disk of a star, giant planets
are extremely hot due to the gravitational potential energy released
in their accretion, and therefore glow brightly at infrared wavelengths.  
Since they have no internal source of energy, they
cool over time, and their infrared flux decreases and shifts toward
longer wavelengths.  Theoretical models \cite{Baraffe} show that the current generation 
of large astronomical telescopes with adaptive optics (AO) systems is 
capable of directly imaging giant extrasolar planets in the infrared.
The cooling properties of extrasolar planets, and the distribution of
stars in the sun's vicinity, suggest two good strategies.  First, a
planet search may focus on very young stars using the near
infrared H and K bands (1.6 and 2.2 microns, respectively).  Stars with ages in
the range of 5-30 million years (Myr) are good targets for this method,
because their giant planets will still be bright at short wavelengths.
Such searches can detect planetary systems that are still in the process
of formation -- a fascinating possibility.  However, stars
that are sufficiently young are rare, and therefore tend to be distant
from the sun, so only bright planets at relatively large physical
seperations can be detected.  A second promising method is to
using longer infrared wavelengths, such as the L' and M bands (3.8 and 4.8 microns, respectively)
to observe nearer stars at more moderate ages.  At these wavelengths,
giant planets can be seen around very nearby stars up to the
5 billion year (Gyr) age of our own solar system.  Younger stars
are still preferred, but excellent sensitivity can be obtained
for stars in the 100 Myr to 1 Gyr range, far older than the
optimal range for H and K searches.  Stars at these ages are
far more common than those younger than 100 Myr, and many examples
can be found very close to the sun.  Very faint planets can be
detected around such nearby stars, and planets at small physical
seperations from their parent stars can be resolved.

We have begun an L' and M band survey of 50 nearby, moderate age
stars for extrasolar planets, using the newly developed L' and M band
camera Clio on the 6.5 meter MMT telescope of the University of Arizona,
with its deformable secondary AO system.  We obtain slightly better sensitivity
to low-mass planets in the L' band than in the M band, so we plan
to do our initial observations in L' and use the M band for followup as needed.
To date, we have observed 7 of our 50 survey targets.  We present here
the methods we use to obtain extremely high contrast and sensitivity.
In Section~\ref{sect:obs} we describe our observations.  Section~\ref{sect:data}
presents our data analysis methods.  In Section \ref{sect:sens} we give
a detailed analysis of the sensitivity we have obtained for planets
orbiting the star GJ 450, one of the 7 targets we have observed so far.
No companions to GJ 450 have been detected.  In Section \ref{sect:mstar}
we announce the discovery of a probable low-mass star orbiting another
of our survey targets, HD 133002.  Section \ref{sect:concl} gives our conclusions.


\section{Observations} 
\label{sect:obs}

On the nights of April 10, 11, and 12, 2006, we obtained long L' band integrations
on 7 nearby, moderately young stars using the Clio camera with the AO system
on the MMT, obtaining an hour of integration or more on every target except 
one, on which we obtained only 30 minutes.  Each integration was composed
of tens to hundreds of individual frames, and each frame was a coadded stack
of either 10 or 25 individual 2 second exposures.  We turned the instrument rotator
off, so that the Clio instrument remained fixed with respect to the telescope.
Because the MMT is on an altazimuth mount, as it tracks celestial objects
they rotate with respect to both the telescope and the instrument.  
Since our frame times were so short, having the instrument rotator off
did not cause any blurring, and the rotation of the celestial objects with respect
to the telescope and instrument was an important part of our overall
strategy (see Section \ref{sect:data}).  We observed
each target as close to its transit as possible, to maximize this rotation as
well as to obtain observations at the lowest airmass.

In order to subtract off glows and other artifacts due to the instrument
and telescope, we took our data in nodded pairs.  That is, we would take
a series (usually 5) frames with the star on one position on the detector,
and then move the telescope about 5-8 arcseconds and take another series
of images with the star in a different detector position.  Then the subtraction
of an image in one nod position from an image in another yeilds an image
in which all telescope and instrument artifacts common to both nod positions
vanish, and any celestial objects remain.  The only disadvantage is the
negative image of the star from the other nod position, behind which, of
course, nothing can be detected.  It is worth this loss of sensitivity
in a small area, however, to ensure that images from both nod positions
can be used for science, and to achieve this end we always arranged the
nod so that the star was well placed on the detector in both nod positions.

With 2 second exposures, the core of the star image was always saturated.  In
order to obtain an estimate of the full point spread function (PSF) obtained
on each target, acquire L' photometry, and monitor the atmospheric transparency,
we took 8-60 shorter frames on each target, interspersed among
the longer exposures.  We used 0.36 or 0.16 second exposure times for these
images to avoid saturating the image cores.

The weather was clear during all of our science observations.  The seeing, however,
was sometimes worse than 1-1.5 arcseconds, and this combined with windshake to
broaden our psf core somewhat beyond the nominal diffraction limit in some cases.  
This reduced the sensitivity slightly, but the data remain very good and suitable for the detection
of extrasolar planets.

\section{Data processing} \label{sect:data}

The basics of our processing pipeline are dark subtraction; flat
fielding; subtraction of nodded pairs of images; 
noise removal through several types of bad pixel fixing and
artifact correction; shifting and rotation of images to move the star to a consistent position on
every image, and to remove the effect of the rotation of
celestial objects with respect to the instrument; and finally
stacking of all the images to produce a master image on which
any faint sources can be detected.  We produce the master image
using a creeping mean combine; that is, if we have n images
we therefore have n values for each pixel.  For each given pixel,
we find the mean of these n values, reject the value farthest from
this mean, find the mean of the remainder, reject the value farthest
from this mean, etc, until a pre-determined fraction of the values,
usually 20 \%, have been rejected.  This powerfully eliminates
artifacts, but real sources, present on every image, are not affected.
As our final step in image processing we apply an unsharp mask
using a gaussian kernal with $\sigma$ = 5.0 pixels, to allow point
sources to stand out above the stellar halo.

In the processing, the fact that celestial objects rotate with respect to the
telescope and instrument during observations becomes important.  Ghosts, rays of scattered
light, and other artifacts present in Clio do not rotate with real celestial
objects, rather, they remain fixed with respect to the telescope and instrument.
Therefore, when the images are rotated in processing to follow real celestial
objects, the ghosts are in a different location on each image and
vanish on the creeping mean combine.  Real sources are not affected.

\begin{figure}
   \begin{center}
   \begin{tabular}{c}
   \includegraphics[height=6cm]{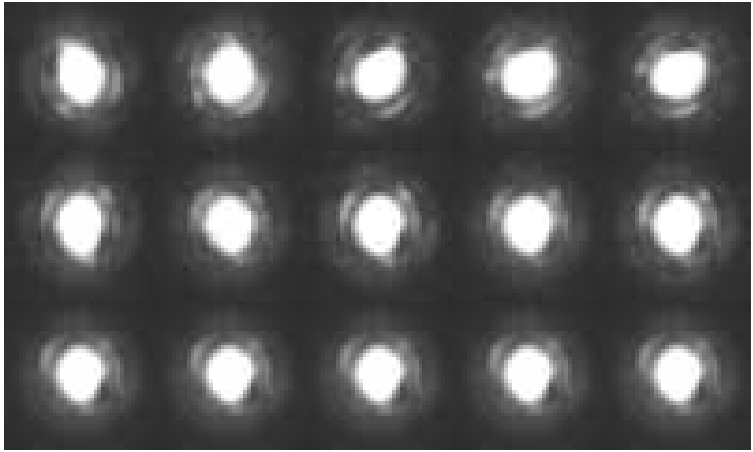}
   \end{tabular}
   \end{center}
   \caption[example] 
   { \label{fig:psftile} 
Images showing how the PSF rotates with respect to real celestial objects, but
remains fixed with respect to the telescope.  The upper row
shows images that have been rotated in processing to be fixed
with respect to any real sources.  The second row shows the same images
fixed with respect to the telescope.  The bottom row simply
shows 5 identical copies of the master PSF image that was
constructed for this data set (see Section \ref{sect:data}), fixed with respect to the telescope.}
   \end{figure}

\begin{figure}
   \begin{center}
   \begin{tabular}{c}
   \includegraphics[height=8cm]{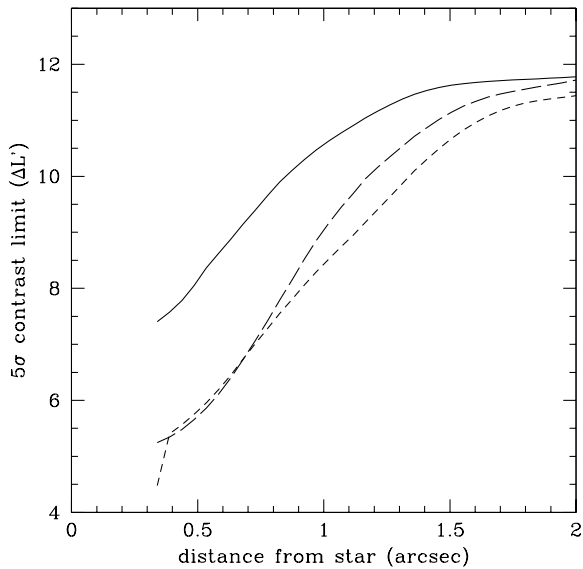}
   \end{tabular}
   \end{center}
   \caption[example] 
   { \label{fig:radsens} 
Highest detectable contrast ratio between the star and the planet in
magnitudes, as a function of radius in arcseconds, for three different
data processing schemes.  The short-dashed line is for our basic
processing with unsharp masking.  The long-dashed line is for
radial arc subtraction, and the solid line is for full psf subtraction
(see Section \ref{sect:data}).  Note that we obtain 5$\sigma$ sensitivity above 10.0 magnitudes,
or a factor of $10^4$, at 1.0 arcsecond seperation with optimal processing.}
   \end{figure}

An additional step in our processing makes further use of the rotation of
real objects with respect to the telescope and instrument.
All AO systems on large telescopes are plagued with 'super-speckles,' that is,
diffraction speckles due to imperfections in the telescope optical surfaces
that are not corrected with AO and which, unlike residual atmospheric speckles,
change only slowly with time and do not smooth out in long integrations.
However, the super speckles remain fixed with respect to the telescope,
while real sources rotate.  Thus, a stack of images fixed
with respect to the telescope may be constructed from a night's data on
a particular star, and these may then be creeping-mean combined with
high rejection, say 50 \%, to create a master PSF image in which only
super speckles and artifacts, no real sources, appear.  This image
may then be subtracted from each individual data frame just prior
to rotation in the ordinary processing sequence.  Superspeckles
are strongly supressed, but real sources are unaffected.  See Figure~\ref{fig:psftile}
for a demonstration of superspeckles and image rotation, and Figure~\ref{fig:radsens}
for the improvement in sensitivity that results from this
processing method.  Previous AO observers have used a similar method\cite{Marois}.

\section{Sensitivity Analysis} \label{sect:sens}
After examining our 7 targets, we decided that the
star GJ 450 would provide the best example of the
typical sensitivity we can obtain.  Our science data
on this star consists of a 5355 second integration made on the night of April 11, 2006.
We analyzed our sensitivity first by obtaining the
approximate PSF of any real sources from
the unsaturated, short exposure images described in
Section \ref{sect:obs}.  We normalized this PSF and
then performed a least-square fit for every pixel in our fully processed combined science images,
to produce a new image containing the integrated flux
of the best-fit PSF centered on each pixel.
Our least-squares fit to the PSF template allowed for
a variable sky background 'pedestal' under the PSF
at each location.  The full width at half maximum (FWHM) of our PSF was typically 
about 3.0 pixels, and we used a fitting radius of 6.0 pixels 
to ensure that the background level was accurately fit.
We then took the image containing the integrated flux
of the best-fit PSF at each pixel, and used it to
construct an estimate of our 5 $\sigma$ sensitivity limit
at each pixel by calculating the rms in a small region
surrounding that pixel and then multiplying it by 5.
For pixels beyond the halo of the primary star we calculated
the rms within a circle of 7-pixel radius, but for pixels
within the primary star halo we used instead a 45 pixel long
arc at fixed radius from the primary star, centered on the
pixel under consideration.
We smoothed the resulting sensitivity image with a gaussian
kernal of $\sigma$ = 3.0 pix, and radially averaged it to
produce Figure~\ref{fig:radsens}, which presents the
5 $\sigma$ contrast limit for companions at different
distances from the star, in the case of our standard
processing, an intermediate processing using the
subtraction of radial arc-averages, and the full
PSF subtraction described in Section \ref{sect:data}.

There are so many unknowns in planet detection, so many
faint artifacts or noise bursts that can imitate real
sources, that no sensitivity analysis is complete
without a Monte-Carlo simulation demonstrating ability
to detect simulated planets added into real data down to a given threshold.
To carry out a Monte-Carlo simulation, we first made an image giving the best 
5 $\sigma$ sensitivity limit obtained by any of the three processing
methods shown in Figure \ref{fig:radsens}, plus one other which
simply involved unsharp masking each individual image before stacking,
and smoothed this image as described above.  Then, we
wrote a code to use this master sensitivity image to insert 
a random selection of planets of fixed significance level into the raw data.  The number,
seperations, and position angles of all these planets were
randomly chosen within user-specified bounds, and their
characteristics were written to a file.  We made data
sets with simulated planets at significance levels of 10 and 5
$\sigma$.  Then, without looking at the file containing the planet characteristics, and thus
without knowing even the number of planets that
had been inserted, we processed the data and tried to
identify the inserted planets on the final images.  It
was immediately obvious that the 5 $\sigma$ planets could
not be confidently detected, and thus our 5 $\sigma$
sensitivity estimates do not correspond to upper
limits on real sources.  

\begin{figure}
   \begin{center}
   \begin{tabular}{c}
   \includegraphics[height=12cm]{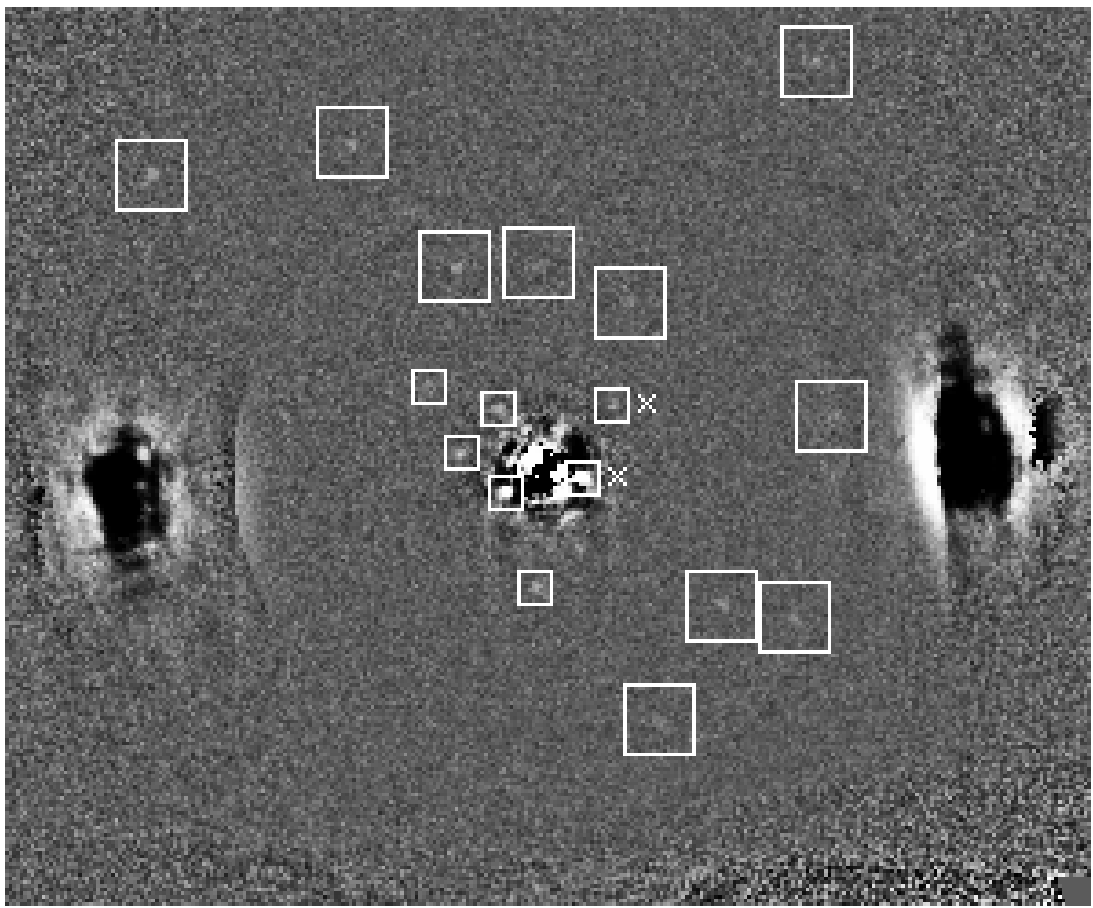}
   \end{tabular}
   \end{center}
   \caption[example] 
   { \label{fig:image} 
Image of GJ 450 with the 17 simulated 10 $\sigma$ planets of the
Monte Carlo simulation inserted.  This is a master image made
from the entire data set with full PSF subtraction and 20 \%
creeping mean rejection in the final combine.  Large boxes are
drawn around the 10 planets that could be confindently detected
without PSF subtraction, while small boxes indicate the 7 that
required subtraction.  Small X's on the right hand side mark
the two boxes showing companions that were not detected
by the automated code, but were, however, correctly identified
manually while still in the blind phase of the simulation.  Dark
regions to the right and left of the central star are residual
negative images from the nod-subtraction.}
   \end{figure}

\begin{figure}
   \begin{center}
   \begin{tabular}{c}
   \includegraphics[height=8cm]{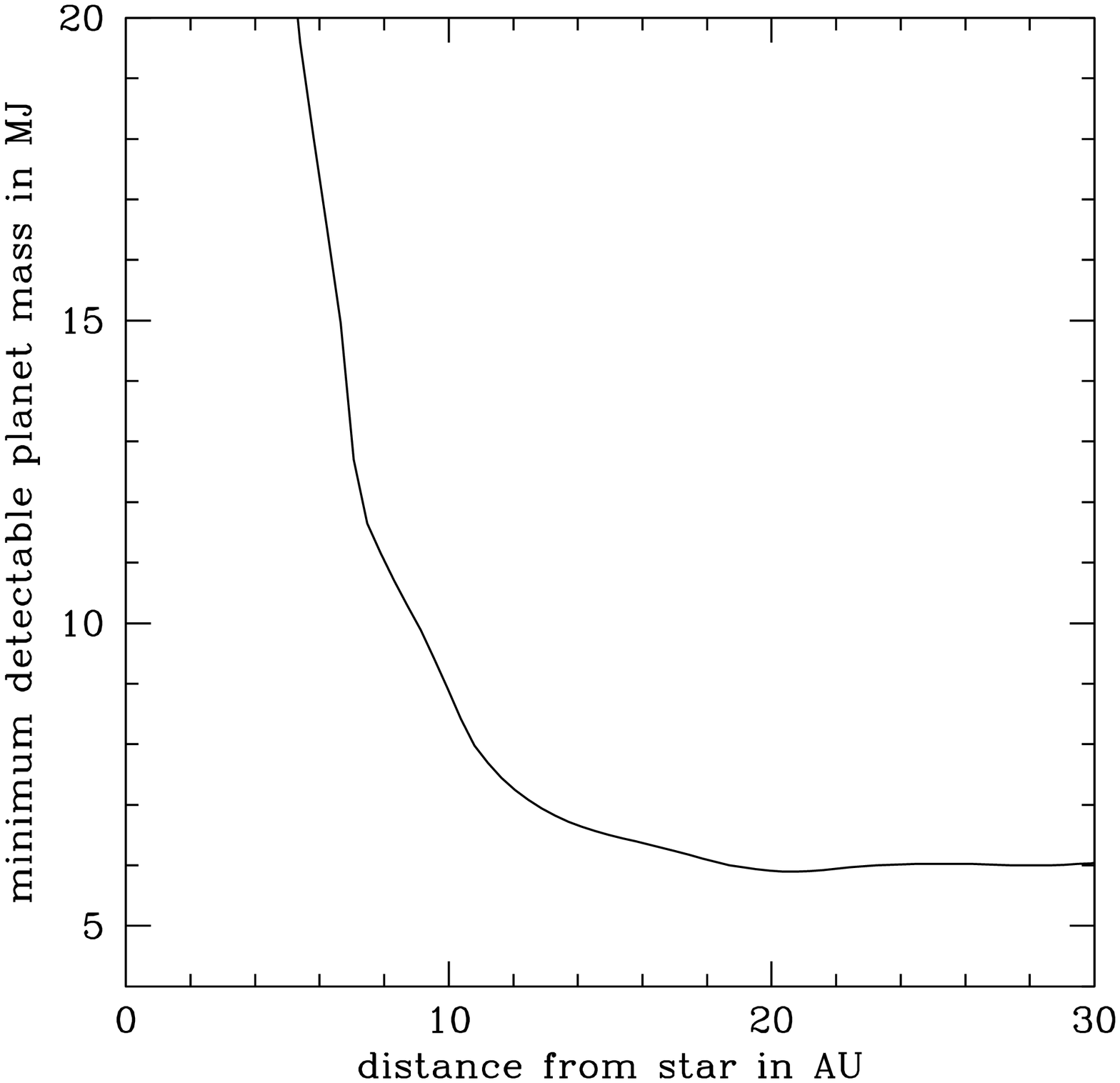}
   \end{tabular}
   \end{center}
   \caption[example] 
   { \label{fig:masssens} 
Mass sensitivity plot, showing the mass of the least massive planet
we could detect, in units of Jupiter masses, vs projected physical
seperation from the star.  These are realistic, 10 $\sigma$ limits,
as confirmed by our Monte Carlo simulation.}
   \end{figure}

We then tried the images with 10 $\sigma$ planets.  Figure
\ref{fig:image} shows a final processed version of these
images, with full PSF subtraction applied.  Several planets
are immediately apparent to the eye, but others could
not be confidently distinguished from artifacts.  We wrote
an automated planet detection code based on the method
by which we constructed the sensitivity images.  A key additional
component of this code was that it did not report the detection of a
source unless it was found on a master image made from all the science data
\textit{and} (albeit at lower significance) on each of two independent master
images constructed from the first and second halves of the
science data.  This requirement was absolutely crucial in
order to avoid tens to hundreds of spurious detections.  
We detected planets on the 10 $\sigma$ images using this automated detection,
and simple examination by eye.  Examination by eye allowed us to rule out a few
of the automated detections, and to include two other sources
which appeared real but had not been automatically detected.
Having thus constructed a master list of apparent planets,
we looked at the 'secret' file written by our simulated
planet code for the first time.  We found that 17 simulated
planets had been inserted into the data at a 10 $\sigma$ sensitivity limit.
We had found them all, and had made no spurious detections.
Our thorough, blind Monte Carlo simulation confirmed our
ability to detect 10 $\sigma$ planets with very high completeness.
Figure \ref{fig:masssens} shows the 10 $\sigma$ mass limits
for planets orbiting GJ 450, as a function of seperation in AU,
based on theoretical models\cite{Baraffe}.  Table \ref{tab:mcs}
gives the data on our 17 simulated planets, including their
masses again based on theoretical models\cite{Baraffe}.  These mass sensitivities
assume an age of 1 Gyr for GJ 450, which is reasonable based
on its X-ray brightness.

\begin{table}[h]
\caption{Properties of simulated 10 $\sigma$ planets used in our Monte Carlo
simulation.
Astrometric error refers to the difference between the true position of the
planet as placed by the simulation code, and the position as derived by
centroiding on the science images.  The errors are impressively low,
indicating proper motion confirmation on any real companions we detect
can be achieved quickly.  Photometric error refers to the true magnitude
of the input planet minus that derived from aperture photometry on the
science image, with an aperture correction from the unsaturated PSF image.
The large negative errors of close-in planets are due to their being
partially subtracted away in the PSF subtraction, which is possible because
of the particular form of the sky rotation vs time for this star.  Our processing
will be modified to reduce or eliminate this effect.} 
\label{tab:mcs}
\begin{center}       
\begin{tabular}{|c|c|c|c|c|} 
\hline
\rule[-1ex]{0pt}{3.5ex}  Seperation & Magnitude & Astrometric error & Photometric error & Mass \\
\rule[-1ex]{0pt}{3.5ex}  (arcseconds) & (L') & (milliarcseconds) & (magnitudes) & (M Jup)\\\hline

\rule[-1ex]{0pt}{3.5ex}  0.51 & 12.53 & 6.99 & -1.17 & 28.08 \\
\hline
\rule[-1ex]{0pt}{3.5ex}  0.56 & 13.32 & 23.85 & -1.07 & 20.55 \\
\hline
\rule[-1ex]{0pt}{3.5ex}  0.95 & 15.35 & 45.53 & -0.73 & 9.85 \\
\hline
\rule[-1ex]{0pt}{3.5ex}  1.14 & 15.6 & 6.94 & -0.46 & 8.96 \\
\hline
\rule[-1ex]{0pt}{3.5ex}  1.27 & 15.96 & 7.97 & -0.83 & 7.66 \\
\hline
\rule[-1ex]{0pt}{3.5ex}  1.58 & 16.06 & 33.25 & -0.18 & 7.40 \\
\hline
\rule[-1ex]{0pt}{3.5ex}  1.90 & 16.51 & 12.51 & -0.28 & 6.05 \\
\hline
\rule[-1ex]{0pt}{3.5ex}  2.50 & 16.59 & 18.57 & -0.38 & 5.89 \\
\hline
\rule[-1ex]{0pt}{3.5ex}  2.69 & 16.57 & 33.44 & -0.42 & 5.91 \\
\hline
\rule[-1ex]{0pt}{3.5ex}  2.91 & 16.38 & 12.7 & 0.12 & 6.44 \\
\hline
\rule[-1ex]{0pt}{3.5ex}  2.98 & 16.6 & 12.89 & 0.12 & 5.87 \\
\hline
\rule[-1ex]{0pt}{3.5ex}  3.71 & 16.51 & 18.71 & -0.17 & 6.05 \\
\hline
\rule[-1ex]{0pt}{3.5ex}  3.90 & 16.59 & 29.55 & -0.28 & 5.88 \\
\hline
\rule[-1ex]{0pt}{3.5ex}  3.93 & 16.62 & 8.64 & -0.72 & 5.83 \\
\hline
\rule[-1ex]{0pt}{3.5ex}  5.02 & 16.49 & 19.78 & -0.05 & 6.11 \\
\hline
\rule[-1ex]{0pt}{3.5ex}  6.52 & 16.43 & 23.21 & -0.22 & 6.29 \\
\hline
\rule[-1ex]{0pt}{3.5ex}  6.53 & 16.27 & 27.08 & 0.1 & 6.78 \\
\hline

\end{tabular}
\end{center}
\end{table}

\section{HD 133002} \label{sect:mstar}

On the night of April 12, 2006, we obtained a 3500 second L' band integration
on the star HD 133002, one of our survey targets.  We detected a faint stellar
companion at seperation $1.8570 \pm 0.0037 ~\mathrm{arcsec}$, position angle
$118.16 \pm 0.15^{\mathrm{o}} $, and L' = $10.87 \pm 0.10$, where approximate
1 $\sigma$ errors are given.

Our careful internet and literature searches have as yet not turned up
any previous mention of this companion, so it appears we have discovered it.
Ironically, our searches did indicate that the star HD 133002 should not
have been on our target list.  We had used the Gliese Catalog parallax for this
star, which indicated a distance of 15.38 pc.  It now appears that the 
Gliese parallax is in error, and the true distance to the star
from Hipparcos is 43.3 pc.  The age also appears to be much larger
than we had thought, as high resolution spectroscopy indicates the star is a metal-poor subgiant \cite{Galeev}.
Simbad records its spectral type as F9.  Given our observed
L' magnitude of $10.87 \pm 0.10$ for the companion, and the Hipparcos distance to the system,
we find that if the companion is physical its absolute L' magnitude is $7.69 \pm 0.10$.  
An M5 V star has an absolute L' magnitude
of 5.71 and a mass of 0.21 solar masses \cite{AAQ}.  The absolute L' magnitude 
of a star of 0.1 solar masses and and age of 1 to 5 Gyr, between
which age limits the HD 133002 system probably lies, is about 9.07 \cite{Baraffe}.  Thus
the companion mass appears to be between 0.1 and 0.21 solar masses,
again assuming the two stars to be physically associated.
Assuming a mass for HD 133002 A of 1 solar masses, and
that the companion is in a circular orbit at the projected
distance, its period is about 700 years.  The accuracy of our
position and the high proper motion of the system imply
a 5 $\sigma$ proper motion confirmation that the companion is
physical will be possible with another measurement of similar
accuracy by early June 2006, and orbital
motion will likely be detectable in 1-2 years.

\section{Conclusion} \label{sect:concl}

The combination of the Clio L' and M band camera and the MMT AO
system is capable of sensitive, high contrast observations in
the L' band.  Detection of planets in the L' band requires extensive 
data processing.  Our method of using sky rotation to remove
super speckles is powerful, at least in the specific case of
the MMT and its AO system.  Estimating the sensitivity of L'
observations in extrasolar planet searches is tricky and very important
in order for upper limits to be meaningful.  Simple sensitivity 
estimates based on the rms
of image regions should never be trusted without confirmation
by Monte Carlo simulation, preferably a carefully constructed
blind simulation such as we have carried out.
Our simulation demonstrated our ability to detect planets
of 6 Jupiter masses orbiting a 1 Gyr old star, and if planets
near this mass are sufficiently common in 10-40 AU orbits we
will probably discover some of them in the course of our survey.
Our method of splitting the data into two independent halves
and requiring that any real source be detectable on both halves
is a good one in order to sort through numerous spurious detections
and find those that are real.  The acqusition of a small
number of short, unsaturated exposures of our science targets
also played an important role in our sensitivity calculation
and our Monte Carlo simulation, as it allowed us to accurately
know the true PSF a planet would be expected to have.

We have discovered a stellar companion to HD 133002.  If it
is physically related to HD 133002, it is a low mass red dwarf
between 0.1 and 0.21 solar masses.  It awaits proper motion
confirmation and other followup studies.

\acknowledgments     
 
We thank Andy Breuninger for many hours spent writing, improving,
and testing the software that runs the Clio camera.  We also thank MMT
telescope operators John McAfee and Michael Alegria, and AO system
operators Doug Miller and Matt Kenworthy for effectively operating
their complex and sometimes finicky systems, and working with us to achieve our
challenging science goals.  This research has made use of the SIMBAD database,
operated at CDS, Strasbourg, France.  Our code for the analyses
described here makes extensive use of Numerical
Recipes\cite{NRC} subroutines.


\bibliography{cliobib}   

\begin{thebibliography}{1}

\bibitem{Baraffe}
I.~Baraffe, G.~Chabrier, T.~S. Barman, F.~Allard, and P.~H. Hauschildt,
  ``Evolutionary models for cool brown dwarfs and extrasolar giant planets. the
  case of hd 209458,'' {\em Astronomy and Astrophysics}~{\bf 402},
  pp.~701--712, 2003.

\bibitem{Marois}
C.~Marois, D.~Lafreni\`{e}re, R.~Doyon, B.~Macintosh, and D.~Nadeau, ``Angular
  differential imaging: A powerful high-contrast imaging technique,'' {\em The
  Astrophysical Journal}~{\bf 641}, pp.~556--564, 2006.

\bibitem{Galeev}
A.~I. Galeev, I.~F. Bikmaev, F.~A. Musaev, and G.~A. Galazutdinov, ``Chemical
  composition of 15 photometric analogs of the sun,'' {\em Astronomy
  Reports}~{\bf 48, 6}, pp.~492--510, 2004.

\bibitem{AAQ}
A.~N.~C. (editor), {\em Allen's Astrophysical Quantities}, Springer-Verlag, New
  York, 2000 (fourth edition).

\bibitem{NRC}
W.~H. Press, S.~A. Teukolsky, W.~T. Vetterling, and B.~P. Flannery, {\em
  Numerical Recipes in C}, Cambridge University Press, New York, 1992 (second
  edition).

\end{thebibliography}
\bibliographystyle{spiebib}   

\end{document}